\documentclass[a4paper,usenatbib]{mn2e}

\usepackage{amsfonts}
\usepackage{amssymb}
\usepackage{amsmath}
\usepackage{graphicx}
\usepackage{color}
\usepackage[totalwidth=505pt,totalheight=680pt]{geometry}

\newcommand{\avg}[1]{\ensuremath{\langle \,#1\, \rangle}}

\newcommand{\eqn}[1]{equation~\eqref{#1}}

\newcommand{\dd}{\mathrm{d}}
\newcommand{\be}{\begin{equation}}
\newcommand{\ee}{\end{equation}}

\newcommand{\PS}{\mathrm{PS}}

\title[Upward mobility and back substitution]
      {The importance of stepping up in the excursion set approach}

\author[M.~Musso, R.~K.~Sheth]
{Marcello Musso$^{1}$\thanks{E-mail: marcello.musso@uclouvain.be} 
 \& Ravi K.~Sheth$^{2,3}$\thanks{E-mail: sheth@ictp.it}\\
 $^1$ CP3-IRMP, Universit\'e Catholique de Louvain, 
      2 Chemin du Cyclotron, 1348 Louvain-la-Neuve, Belgium \\
 $^2$ The Abdus Salam International Center for Theoretical Physics,
      Strada Costiera, 11, Trieste 34151, Italy\\
 $^3$ Center for Particle Cosmology, University of Pennsylvania, 
      209 S. 33rd St., Philadelphia, PA 19104, USA}

\begin{document}

\pagerange{\pageref{firstpage}--\pageref{lastpage}}

\maketitle 

\label{firstpage}

\begin{abstract}
Recently, we provided a simple but accurate formula which closely approximates the first crossing distribution associated with random walks having correlated steps.  The approximation is accurate for the wide range of barrier shapes of current interest and is based on the requirement that, in addition to having the right height, the walk must cross the barrier going upwards.  Therefore, it only requires knowledge of the bivariate distribution of the walk height and slope, and is particularly useful for excursion set models of the massive end of the halo mass function.  However, it diverges at lower masses.  We show how to cure this divergence by using a formulation which requires knowledge of just one other variable.  While our analysis is general, we use examples based on Gaussian initial conditions to illustrate our results. Our formulation, which is simple and fast, yields excellent agreement with the considerably more computationally expensive Monte-Carlo solution of the 
 first crossing distribution, for a wide variety of moving barriers, even at very low masses.
\end{abstract}

\begin{keywords}
large-scale structure of Universe
\end{keywords}

\section{Introduction}
Simulations of hierarchical gravitational clustering suggest that the abundance and clustering of gravitationally bound objects in the Universe can be a powerful tool for constraining the nature of the initial fluctuation field.  Since simulations are expensive, there is considerable interest in models which can provide a better understanding of how cluster abundances and clustering depend on cosmological parameters.  The excursion set approach \citep{bcek91} is perhaps the most developed of these:  motivated by the seminal work of \cite{ps74} it provides an analytical framework which relates the statistics of gravitationally bound dark matter haloes to fluctuations in the primordial density field, and the subsequent expansion history. 

In this approach, at a given (randomly chosen) position in space one looks at the overdensity field smoothed on some scale $R$: plotting this smoothed $\delta$ as a function of (the inverse of) $R$ resembles a random trajectory, the steps of which are, in general, correlated.  The nature of the correlations depends on the smoothing filter (e.g. tophat, Gaussian), and on the nature of the initial fluctuation field (Gaussian or non-Gaussian).  Repeating this for every position in space gives an ensemble of trajectories, each one of which starts from $\delta(R=\infty)=0$ (the universe is homogeneous on large smoothing scales).  For each trajectory, one searches for the largest $R$ (if any) for which the value of the smoothed density field lies above some threshold value (which may itself depend on $R$), the value of which is determined by the expansion history of the background cosmology.  An object of mass $M\sim  R^3$ is then associated with that trajectory.  

If $\dd n/\dd M$ denotes the comoving number density of haloes of mass $M$, then the mass fraction in such halos is $(M/\bar\rho)\,\dd n/\dd M$, where $\bar\rho$ is the comoving background density.  The excursion set approach assumes that this halo mass fraction equals the fraction of walks which cross the threshold (the ``barrier'') for the first time when the smoothing scale is $R$:
\begin{equation}
 f(R)\,\dd R = (M/\bar\rho)\,(\dd n/\dd M)\, \dd M.
 \label{ansatz}
\end{equation}
Although recent work has focussed on the shortcomings of this ansatz \citep{ps12}, not to mention the fact that variables other than the overdensity affect halo formation, and this is not evident in the simplest version of the approach outlined above \citep{smt01}, the first crossing distribution is nevertheless expected to provide substantial insight into the dependence of $\dd n/\dd M$ on cosmological parameters.  

In practice, one works not with $f(R)$ but with $f(s)$, where 
\begin{equation}
 s(R) \equiv \int \frac{\dd k}{k}\, \frac{k^3P(k)}{2\pi^2}\,W^2(kR)
\end{equation} 
denotes the variance in the fluctuation field when smoothed on scale $R$ with a filter of shape $W$.  In hierarchical models, $s$ is a monotonic function of $R$, so $f(R)\dd R = f(s)\dd s$.  Working with $s$ has the advantage of removing most of the dependence on the shape of the power spectrum:  $P(k)$ mainly matters only through the dependence of $s$ on $R$.  

To solve the first crossing problem, we must be able to identify the fraction of trajectories for which $\delta = b(s)$ and $\delta < b(S)$ for all $S < s$.  While this is rather straightforward to implement numerically \citep{bcek91}, the latter condition is hard to deal with analytically.  Recently, \citet{ms12} argued that a good approximation to $f(s)$ can be computed from considering the joint probability $p(\delta,v;s)$ that a walk reaches $\delta$ at scale $s$ with velocity $v \equiv \dd\delta/\dd s$.  Namely, they argued that if $\delta = b(s)$, then one also wants $v \ge \dd b/\dd s$, so 
\begin{equation}
  f(s) \simeq p(b) \!\int_{b'}^{\infty} \!\!\mathrm{d}v \, (v-b') \, p(v|b) \,.
 \label{fms}
\end{equation}

For Gaussian statistics, $p(b)$ is a Gaussian with zero mean and variance $s$, and $p(v|b)$ is Gaussian with mean $\avg{\!v|b\!} = b\,\avg{\!\delta v\!}/\avg{\!\delta\delta\!}=b/2s$ and variance $\avg{\!vv\!} - \avg{\!\delta v\!}^2\!/\avg{\!\delta\delta\!}=s\avg{\!(\delta/\!\sqrt{s})'^2\!}$; this makes 
\begin{equation}
 sf_{\rm MS}(s) = sf_{\rm PS}(s)
% f(s) = %\left(\frac{b}{2s} - \frac{\dd b}{\dd s}\right)\,
%        -\frac{\dd (b/\sqrt{s})}{\dd s}\,\frac{{\rm e}^{-b^2/2s}}{\sqrt{2\pi}}
  \bigg[\frac{1 + {\rm erf}(x/\!\sqrt{2})}{2} 
       + \frac{{\rm e}^{-x^2/2}}{\sqrt{2\pi}\,x}\bigg]\,,
% \label{fmsg}
\label{sfms}
\end{equation}
where 
\begin{equation}
  x \equiv -\frac{ (b/\sqrt{s})'}{\avg{\!(\delta/\sqrt{s})'^2\!}^{1/2}}
  = -2\Gamma s\, \frac{\dd (b/\!\sqrt{s})}{\dd s}\,,
\end{equation}
with $\Gamma^2 = \gamma^2/(1-\gamma^2)$ and 
$\gamma^2\equiv \avg{\!\delta v\!}^2/\avg{\!\delta\delta\!}\avg{\!vv\!}=1/4s\avg{\!v^2\!}$, and
\begin{equation}
  f_{\rm PS}(s) =
  -\,\frac{\dd (b/\sqrt{s})}{\dd s}\,\frac{{\rm e}^{-b^2/2s}}{\sqrt{2\pi}}
%  = \frac{x}{2\Gamma}\, p(b/\sqrt{s})
\end{equation}
is the old approximation of \cite{ps74}. For future reference, we note that one can also write $f_{\rm PS}(s) = \avg{\!(v-b')|b\!}\, p(b)$, as can be seen assuming $b'\ll\avg{\!v|b\!}$ in \eqn{fms}.  
For a constant barrier $b=\delta_c$, if we define $\nu\equiv\delta_c/\sqrt{s}$, then $\dd (b/\sqrt{s})/\dd s = -\nu/2s$ and $x = \Gamma\nu$, so that one gets the familiar
\begin{equation}
 sf_{\rm PS}(s) = \frac{\nu}{2}\,\frac{{\rm e}^{-\nu^2/2}}{\sqrt{2\pi}}\,.
\label{sfps}
\end{equation}
This highlights the fact that the Musso-Sheth approximation can be thought of as providing a correction to the Press-Schechter result. This correction only matters when $x\lesssim 1$ (or $\nu\lesssim 1/\Gamma$).

This cannot be the full story of course, because this procedure fails to discard those walks that were above threshold at $S<s_1<s$, but diffused back below the barrier at $s_1$ to cross again at $s$, i.e.~ $\delta>b(S)$ but $\delta<b(s_1)$. However, \citet{ms12} pointed out that the fraction of such walks should be tiny at small $s$, since the correlations between the steps make sharp turns very unlikely, and showed that \eqn{fms} indeed works very well for the small values of $s$ (i.e. large $R$ and hence large $M$) which are of most interest in cosmology.  This accuracy is remarkable, since it only requires knowledge of (a judiciously chosen!) bivariate distribution, rather than of the full $n$-point distribution which describes the walk heights at each $S\le s$ (and note that $n\to\infty$).  

The success and simplicity of the approach motivates the search for an approximation that allows even greater accuracy. Since the integral of \eqn{fms} over all $s$ diverges, it is clear that something new is needed at larger $s$ (i.e., smaller $m$).  This is not surprising, since \eqn{fms} fails to discard walks with additional crossings at $S<s$, and these can be a large fraction as $s$ increases. Therefore, we would like to find a scheme that removes this divergence, and to do it by keeping track of as few additional correlated variables as possible.  
We show how to do this is Section~\ref{bs}, where we treat the case of a constant barrier in some detail, before generalizing to moving barriers.  %Section~\ref{toy} describes a toy model which is the next more complicated model to one in which there are no correlations between steps.  We use this model to illustrate why the approach of the previous section works so well.  We also show that, despite its simplicity, this model happens to provide a very good description of the first crossing distribution associated with walks having the full correlation structure associated with $\Lambda$CDM power spectra.  
A final section summarizes our results.  For many aspects of this problem, fully analytic results can be obtained from Gaussian smoothing, i.e. with $W(kR) = \exp(-k^2R^2/2)$, of a Gaussian field with power spectrum $P(k)\propto k^{-1}$; these are collected in Appendix~\ref{n=-1}.  Appendix~\ref{algorithm} gives details of the numerical backsubstitution algorithm we use to derive our results.

%%%%%%%%%%%%%%%%%%%%%%%%%%%%%%%%%%%%%%%%%%%%%%%%%%%%%%%%%%%%
\section{Back-substitution with correlated steps}
\label{bs}
%%%%%%%%%%%%%%%%%%%%%%%%%%%%%%%%%%%%%%%%%%%%%%%%%%%%%%%%%%%%

The starting point of the analysis of \citet{ms12} is that one can write the first crossing distribution as
\begin{equation}
  f(s) = \int_{b'}^\infty \!\!\! \dd v \, (v-b') \,
  p(b,v, \mathrm{first~} s) \,;
\label{fsexact}
\end{equation}
dropping the `first $s$' constraint, while keeping the upcrossing requirement $v>b'$, leads to \eqn{fms}. 
\citeauthor{ms12} showed that this actually works very well down to $\Gamma\nu\gtrsim 0.1$.  Earlier crossings may be accounted for by mean of recursive corrections (see the discussion on how to do this given by \citealp{ms13a}), which introduce two additional variables for each crossing.

One may instead consider the probability $p_{\times\!}(\delta)$ that a walk reaches $\delta$ at scale $s$ having crossed the barrier at least once at larger scale. Following the same approach, and classifying all walks by the scale $S$ on which they first crossed, this can be written -- exactly -- as
\begin{equation}
\label{pcross}
  p_{\times\!}(\delta,s) = \int_0^s\!\dd S \!\int_{B'}^\infty \!\!\! \dd V(V-B') \,
  p(\delta, B,V, \mathrm{first~} S)\,,
\end{equation}
where $B$ and $V\equiv \dd \delta/\dd S$ denote the barrier height and the slope of the walk on scale $S$.
Since all walks that are above the barrier on scale $s$ must have crossed it at some larger scale $S<s$, then $p_{\times\!}(\delta\ge b)=p(\delta\ge b)$ necessarily. Multiplying and dividing the integrand by $f(S)$ leads to
\begin{equation}
 p(\delta\ge b,s) =
  \int_0^s \dd S\, f(S)\,p(\delta\ge b,s|{\rm first}\ S)\,,
 \label{backsub}
\end{equation}
where (considering for simplicity a barrier of constant height $b=B=\delta_c$, for which $B'=0$) we can formally identify
\begin{equation}
  p(\delta\!\ge b,s|\mathrm{first~}S) \equiv \int_0^\infty \!\!\!\dd V\,V\,
  \frac{p(\delta\!\ge b,V,B,\mathrm{first~}S)}{f(S)}
\label{dcondfcS}
\end{equation}
as the conditional probability of a walk ending above threshold at $s$ given that it crossed for the first time at $S$.
%\begin{equation}
%  p(\delta\!\geq\delta_c,s)=
%  \int_0^s\!\dd S \!\int_0^\infty \!\!\dd V\,V\,
%  p(\delta\ge\delta_c,V,B,\mathrm{first~}S)\,,
%\end{equation}
%where, for constant barriers, $b = \delta_c$.   

For a Gaussian field the left-hand side of \eqn{backsub} is just ${\rm erfc}(b/\sqrt{2s})/2$; therefore, if one can come up with a good approximation for $p(\delta\ge b,s|{\rm first}\ S)$, then $f(S)$ can be solved for numerically by simple back-substitution.  Note however that this logic is general, applying as well to both Gaussian and non-Gaussian walks.  Since one usually knows the quantity on the left-hand side, the problem reduces to guessing $p(\delta\ge b,s|{\rm first}\ S)$. Of course, the devil is in this last detail.

%I.e., one can perform the back-substitution with the appropriate non--Gaussian distributions on the left and right-hand sides of equation~(\ref{backsub}).  

\subsection{Separability}
If $p(\delta\ge b,s|{\rm first}\ S)$ is a separable function of $s$ and $S$, then the expression above can be manipulated to obtain a simple expression for $f(s)$.  Namely, if
\begin{equation}
 \label{gShs}
 p(\delta\ge b,s|{\rm first}\ S) = g(S)\,h(s),
\end{equation}
then 
\begin{equation}
 p(\delta\ge b,s) = h(s) \, \int_0^s \dd S\, f(S)\,g(S),
\end{equation}
making 
\begin{equation}
 f(s) = \frac{1}{g(s)}\frac{\dd}{\dd s}
        \bigg[\frac{p(\delta\ge b,s)}{h(s)}\bigg]\,.
 \label{fsgShs}
\end{equation}
For walks with completely correlated or completely uncorrelated steps, and a constant barrier $\delta_c$, the product $g(S)\,h(s)$ equals 1 and $1/2$ respectively \citep{pls12,bcek91}.  Since this is independent of both $s$ and $S$, both limits yield particularly simple relations between the first crossing distribution and the probability distribution of $\delta$ on the first crossing scale:  these limits yield the expressions for $f(s)$ derived by \citet{ps74} and \citet{bcek91} respectively.  

In general, of course, $p(\delta\ge b,s|{\rm first}\ S)$ is more complicated.  However, this simple case is still very instructive because it shows that the back-substitution method can interpolate between these two regimes. It is therefore more general than the approach of \citet{ms12}, which becomes increasingly accurate as the correlation between steps increases, but diverges in the limit of uncorrelated steps.  

\subsection{Normalization}
\label{sec:norm}
Before we study the general case, it is worth noting that this approach is attractive also because it provides an easy way to see that the result will be correctly normalized.  While this is trivial to see geometrically, the argument relies on using the correct expression for $p(\delta\ge b,s|{\rm first}\ S)$.  

For a constant barrier, there is equal probability of ending up on either side of it if one waits long enough: evaluating \eqn{backsub} at $s=\infty$ one has therefore
\begin{equation}
 \frac{1}{2} = \int_0^\infty \dd S\,f(S)\,
                p(\delta\ge b,\infty|{\rm first}\ S)\,.
 \label{normalized}
\end{equation}
For the same reason, it must also be true that $p(\delta\ge b,\infty|{\rm first}\ S)=1/2$, which shows that the exact solution has $\int_0^\infty {\rm d}S\,f(S)=1$.  I.e., $f(S)$ is correctly normalized to unity (except for walks with completely correlated steps, where the integral gives 1/2, as it should). 

More importantly, this relation also ensures that, if the approximate $p(\delta\ge b,s|{\rm first}\ S)$ that we choose has the same large-$s$ limit, then the approximate $f(S)$ that we get from \eqn{backsub} is correctly normalized to unity (for constant barriers; we will come back to this point in the case of moving barriers). Our goal will be to find such an approximation for the conditional distribution.  

\subsection{The simplest approximation}
We remarked earlier that the performance of this approach depends critically on having a good approximation for $p(\delta\ge b,s|{\rm first}\ S)$.  If we require that the walk had height $B$ on scale $S$, but drop the constraint that the barrier had not been crossed prior to this, then one might approximate 
\begin{equation}
 p(\delta\ge b,s|{\rm first}\ S) \approx p(\delta\ge b,s|B,S)\,;
 \label{badApprox}
\end{equation}
the result of inserting this in equation~(\ref{backsub}) can be used to obtain an approximation for $f(S)$ by numerical back-substitution.  

For a Gaussian process and a barrier of constant height, 
\begin{equation}
 p(\delta\ge\delta_c,s|\delta_c,S) = \frac{1}{2}\,
  {\rm erfc}\bigg[\frac{\delta_c(1 - S_\times/S)}{\sqrt{2}(s - S_\times^2/S)^{1/2}}\bigg],
 % {\rm erfc}\left(\frac{\delta_c[1 - S_\times/S]}{\sqrt{2s(1 - (S_\times/S)^2(S/s)}}\right),
\label{badApproxGauss}
\end{equation}
where $S_\times\equiv \langle \delta\Delta\rangle$, and $\Delta$ denotes the walk height on scale $S$.  To see that the resulting estimate for $f(s)$ will be correctly normalized, recall that \eqn{normalized} shows that we must look at the $s\to\infty$ (i.e. the $r\to 0$) limit of this expression.  For the filters of interest here $S_\times/S$ is finite, so this limit is ${\rm erfc}(0)/2 = 1/2$, indicating that the estimate for $f(s)$ which results from approximation~\eqref{badApprox} will be normalized to unity.  

Although this approach (equation~\ref{badApprox} in~\ref{backsub}) has been followed in the past \citep[e.g.][]{j95, n01}, all previous work has failed to appreciate that \eqn{badApprox} is actually only an approximation, and, for most filters of interest, a bad one at that.  To see how bad, we show in the Appendix that this problem can be solved exactly, analytically, for Gaussian smoothing of a Gaussian field with $P(k)\propto k^{-1}$, and that the resulting expression for the first crossing distribution is much further from the correct answer than is $f_\mathrm{MS}$.  

For walks with uncorrelated steps, on the other hand, this approach is exact. For a constant barrier it returns \eqn{fsgShs} with $g=1/2$ and $h=1$. For a moving barrier, \eqn{badApproxGauss} would depend on  $(b-B)/\sqrt{(s-S)}$, while the l.h.s of \eqn{backsub} would be $(1/2)\mathrm{erfc}(b/\sqrt{2s})$.

\subsection{The two-step approximation}
How might one improve on this approximation?  One obvious possibility is to add the additional constraint that the walk height at one other (larger) scale was less than $\delta_c$:
\begin{equation}
 p(\delta\ge\delta_c,s|{\rm first}\ S) \approx 
 \frac{\int_{-\infty}^{\delta_c}{\rm d}\Delta_1
       \int_{\delta_c}^{\infty}{\rm d}\delta\,p(\delta,s,\Delta_1,S_1|\delta_c,S)}
      {\int_{-\infty}^{\delta_c}{\rm d}\Delta_1\,p(\Delta_1,S_1|\delta_c,S)}
 %\frac{\int_{-\infty}^\mu{\rm d}y\,p(y|\eta)\,{\rm erfc}(\mu_{\nu\eta y}/\sqrt{2})/2}{\int_{-\infty}^\mu{\rm d}y\,p(y|\eta)}
 \label{2steps}
\end{equation}
for some $S_1\le S$.  

To see that there is little point in having $S_1\ll S$, first write 
 $p(\delta,\Delta_1|\delta_c,S) = p(\Delta_1|\delta_c,S)\,p(\delta|\Delta_1,\delta_c)$.  These are all Gaussian distributions with different means and variances.  The distribution $p(\delta|\Delta_1,\delta_c)$ only depends on $\Delta_1$ and $\delta_c$ through its mean, which is given by 
 $\avg{\delta|\Delta_1,\delta_c} = r_{c\delta}\delta_c 
  + (r_{1\delta} - r_{c\delta} r_{1c})(\Delta_1 - r_{1c}\delta_c)/(1 - r_{1c}^2)$.
If $S_1\ll S$ then $r_{1c}\ll 1$, and since $r_{1\delta}\le r_{1c}\ll 1$ 
 $\avg{\delta|\Delta_1,\delta_c} \to r_{c\delta}\delta_c$.  Thus, in this limit, $p(\delta|\Delta_1,\delta_c) \to p(\delta|\delta_c)$.  As a result, the integrals in the numerator become separable, and one of them is cancelled by the integral in the denominator, leaving $p(\delta\ge\delta_c,s|\delta_c,S)$.  I.e., when $S_1\ll S$ then this approximation reduces to equation~\eqref{badApprox}.  Thus, this additional constraint only matters when the correlation coefficient between the two scales is close to unity:  i.e., when $S_1\sim S$.  The $S_1\to S$ limit is, essentially, the Musso-Sheth approximation, which we know works very well, especially at large $\nu$.

\subsection{The importance of stepping up}

The success of \citet{ms12} in approximating $f(s)$ suggests that further improvement may be obtained by relaxing the `first $S$' constraint in $p(\delta\ge b,s|\mathrm{first~} S)$ into the simpler requirement of merely upcrossing at $S$. Therefore, \eqn{backsub} may be approximated by
\begin{equation}
  p(\delta\ge b,s) \approx
 \int_0^s \dd S\,f(S) \,p(\delta\ge b,s|\mathrm{up~} S)\,,
\label{approxbacksub}
\end{equation}
where the conditional probability can be obtained dropping `first $S$' in \eqn{dcondfcS}, and therefore replacing $f(S)$ with $f_\mathrm{MS}(S)=\int_0^\infty {\rm d}V\,V\,p(V,B)\,$; for $b=B=\delta_c$ this leads to
\begin{align}  
% p(\delta\ge b(s)|{\rm first}\ S) \approx 
% \frac{\int_{b(s)}^\infty{\rm d}\delta 
%       \int_{B'}^\infty {\rm d}V\,(V-B')\,p(\delta,V|B)}
%      {\int_{B'}^\infty {\rm d}V\, (V-B')\, p(V|B)},
 p(\delta\ge\delta_c,s|\mathrm{up~} S) =
  \frac{\int_0^\infty \!\dd V V\,p(\delta\ge\delta_c,V|B)}{\int_0^\infty {\rm d}V\,V\,p(V|B)}\,,
%   \frac{\int_0^\infty {\rm d}V\,V\,p(\delta\ge\delta_c,V,B)}
%        {\int_0^\infty {\rm d}V\, V\, p(V,B)} \notag \\
% %p(\delta\ge\delta_c,s|{\rm first}\ S) \approx 
% %\frac{\int_0^\infty {\rm d}V\, V p(V|b) \int_{\delta_c}^\infty {\rm d}\delta \, p(\delta|b,V)}{\int_0^\infty {\rm d}V\, V\, p(V|b)},  &\equiv
%  \frac{p_\mathrm{MS}(\delta\ge\delta_c,s|S)}{f_\mathrm{MS}(S)} \,,
 \label{msApprox}
\end{align}
%Although these expressions ignore the constraint that $S$ was the first crossing scale, 
and the upcrossing is guaranteed by the fact that the walk must have positive slope at $S$.  For small enough $S$ (meaning $S\ll \delta_c^2$), most walks of height $\delta_c$ will have positive slope, so this constraint should matter little, but at large $S$ it may begin to play a role.  
Finally, for $s\gg S$ the probability of having $\delta>\delta_c$ is 1/2 regardless of $V$, which ensures that $f(s)$ derived from \eqn{approxbacksub} is correctly normalized.

We have already seen that $p(\delta\ge\delta_c,s|\mathrm{up~} S) \simeq 1$ implies that one recovers $f_\mathrm{PS}$ from \eqn{fsgShs}. This is in fact what happens with correlated steps at small $s$, when, if the walk reaches the barrier at $S<s$, it is almost certainly crossing upwards and thus remains above threshold at $s$. In this regime, both numerator and denominator in \eqn{msApprox} are approximately equal to $\avg{\!V|B\!}=\int_{-\infty}^{+\infty} \!{\rm d}V V\,p(V|B)$.

Similarly, it is instructive to see under what conditions one can recover $f_\mathrm{MS}$. For instance, this is the case if only the numerator equals $\avg{\!V|B\!}$: then, because the denominator does not depend on $s$, the resulting expression will be the product of $h(s)=1$ and $g(S)=f_\mathrm{PS}(S)/f_\mathrm{MS}(S)$, which in \eqn{fsgShs} will yield the Musso-Sheth approximation for $f(S)$.  In fact, this picture is correct at larger $s$, when the fraction of walks crossing $\delta_c$ downwards becomes relevant; the numerator includes values of $\delta>\delta_c$ for which earlier crossings are less and less likely, and thus on average it remains close to $\avg{\!V|B\!}$ for longer than the denominator.  We provide an analysis of the $n=-1$ case showing this explicitly in the Appendix.  

\begin{figure*}
 \centering
 \includegraphics[width=0.33\hsize]{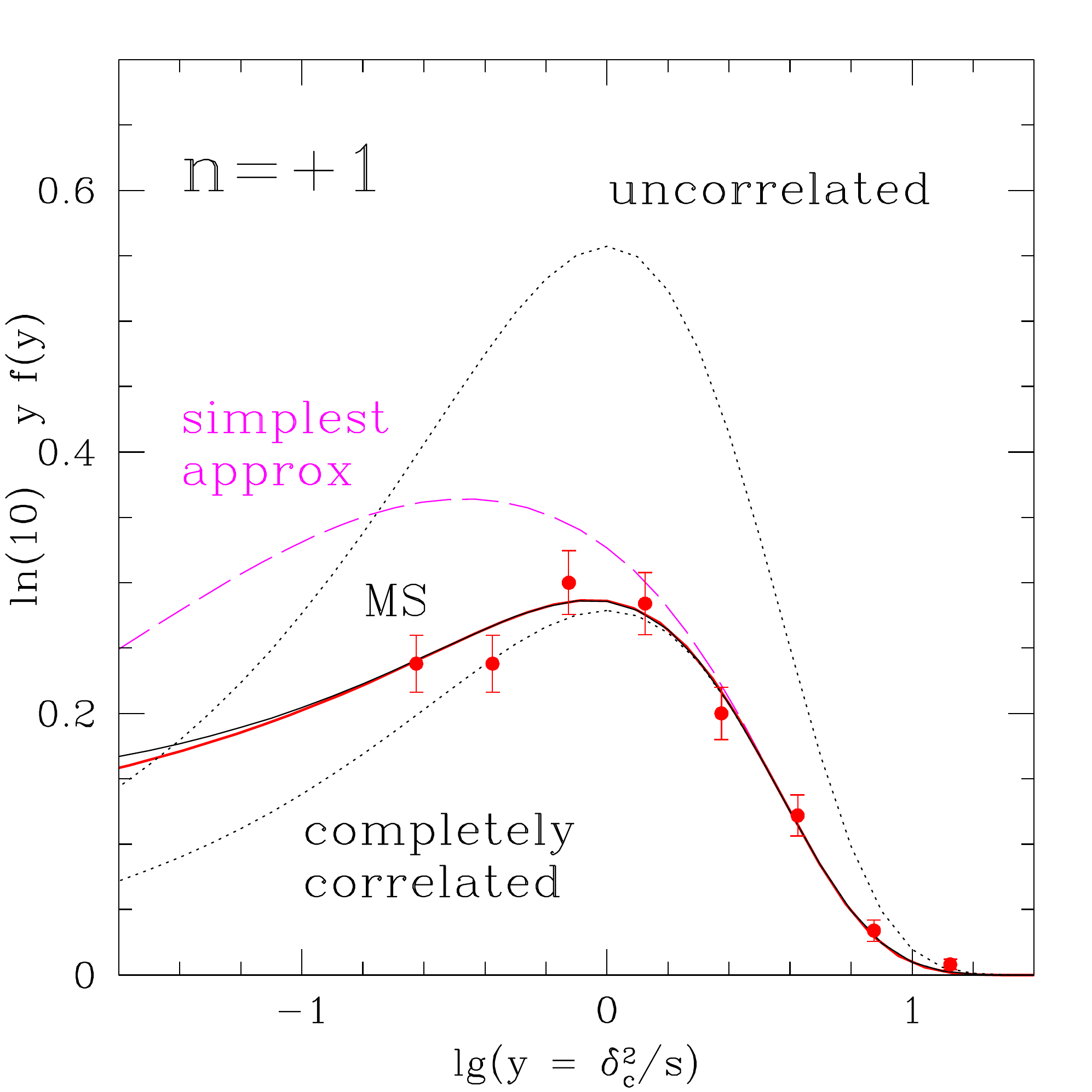}
 \includegraphics[width=0.33\hsize]{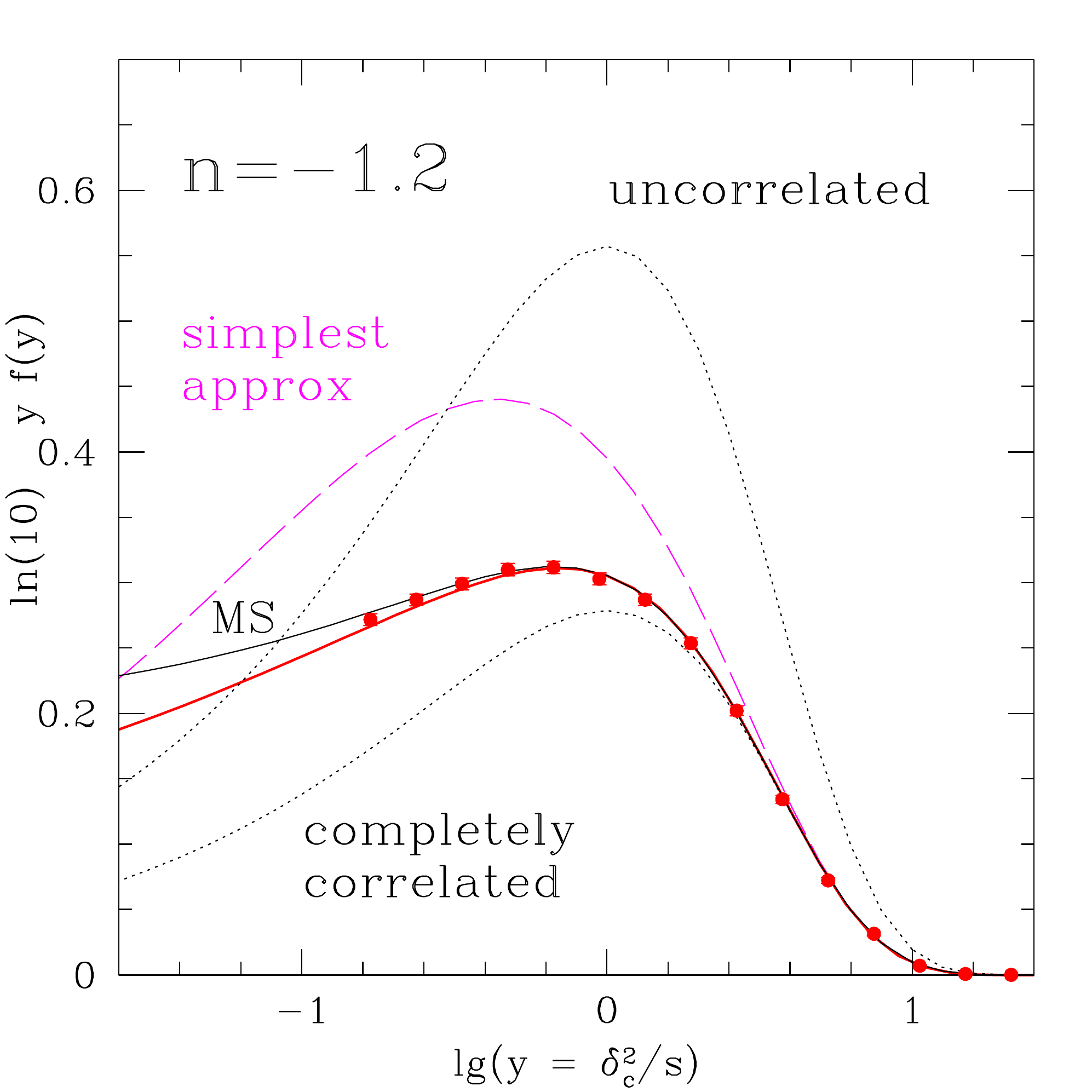}
 \includegraphics[width=0.33\hsize]{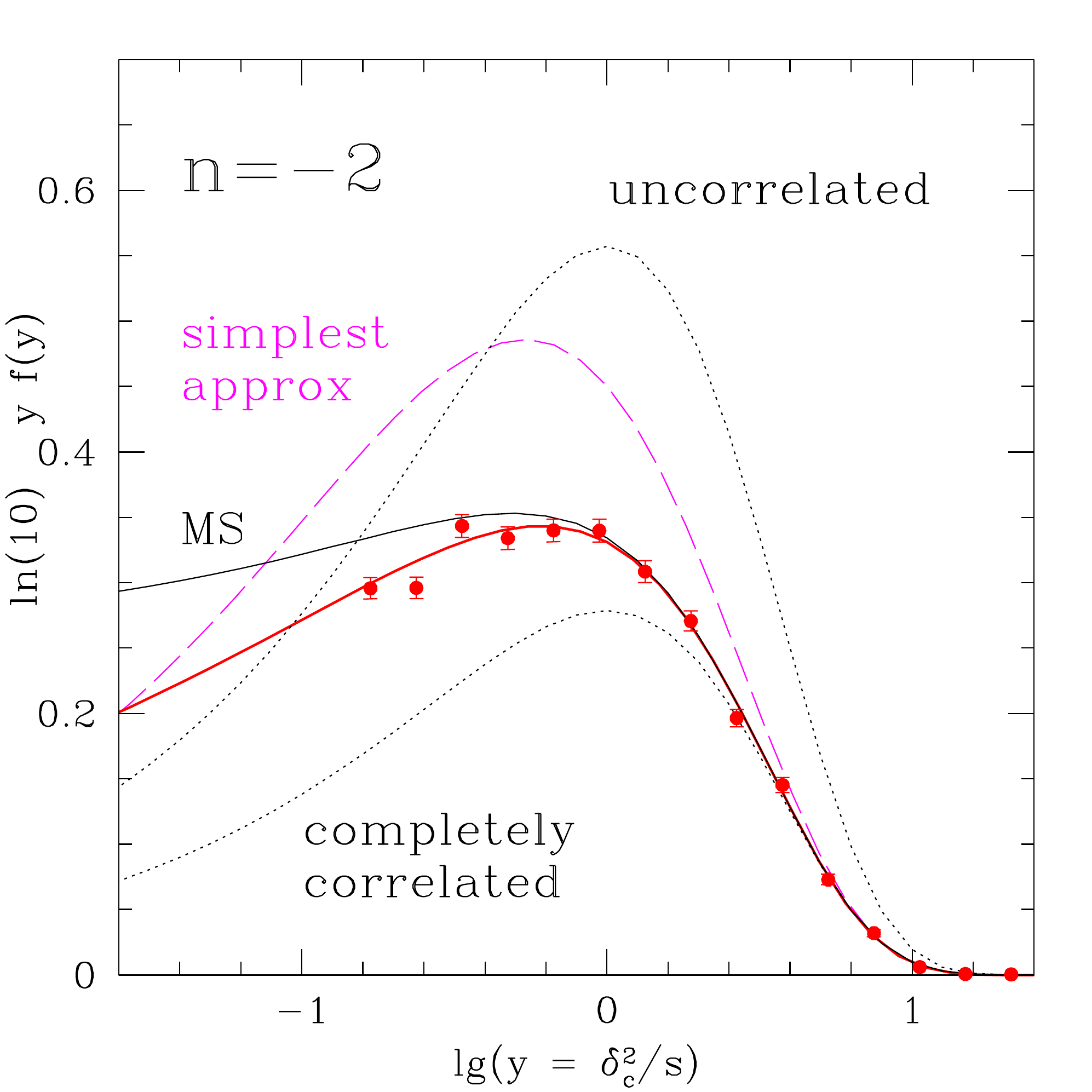}
 \caption{Comparison of various approximations for the distribution of 
          the scale $y=\delta_c^2/s$ on which walks associated with 
          Gaussian smoothing of a power spectrum with $P(k)\propto k^n$ 
          first cross a barrier of constant height $\delta_c$.  
          In each panel, the long dashed curve shows the result 
          of inserting the simple approximation~(\ref{badApprox}) 
          in~\eqn{backsub}; thin solid curve shows the fully analytic 
          approximation of Musso \& Sheth (2012); and thick solid curve 
          shows the better approximation which follows from using 
          \eqn{msApprox} in~\eqn{backsub}.  This last provides an excellent 
          description of the exact solution obtained by Monte-Carlo methods 
          (symbols with error bars).}
 \label{bsall}
\end{figure*}

To recover the approximation of \eqn{badApprox}, suppose (wrongly!) that $p(\delta,V|B) = p(\delta|B)\,p(V|B)$.  Then the integral over $V$ simplify with the denominator, so only the integral over $p(\delta|B)$ remains, and the resulting expression reduces to \eqn{badApprox}.  This shows that \eqn{badApprox} is the result of assuming that the walk height $\delta$ on scale $s$ is only correlated with the slope $d\Delta/dS$ of the walk on a larger scale $S$ because both $\delta$ and $d\Delta/dS$ are correlated with $\Delta$.

To see the effect of the $\delta\,$-$V$ correlation, let us compute \eqn{msApprox} explicitly.  This is most easily done by using the rescaled barriers $\nu\equiv b/\sqrt{s}$ and $\eta\equiv B/\sqrt{S}$, and the unit variance variables that factorize $p(B,V)$, that is $\Delta/\sqrt{S}$ and $u\equiv2\Gamma S(\Delta/\sqrt{S})'$, with $\Gamma = 1/\sqrt{4S\avg{\!\Delta'^2\!}-1}$.  This yields 
\begin{equation}
 p(\delta\ge\delta_c,s|{\rm up}\ S) = 
 \frac{\int_{-X}^\infty{\rm d}u(u+X)\,p(u)\,{\rm erfc}(\mu_{\nu\eta u})}
% {2\avg{x|\eta}_0}
 {2\int_{-X}^\infty{\rm d}u\,(u+X)\,p(u)}\,,
\label{explicit}
\end{equation}
where $X\equiv\Gamma\eta$ and, after defining $\xi\equiv\avg{\!(\delta/\sqrt{s})(\Delta/\sqrt{S})\!}$ and $\Sigma\equiv \avg{\!(\delta/\sqrt{s}) u\!}$,
%$\xi\equiv\avg{\delta\Delta}/\sqrt{sS}$,
\begin{equation}
 \mu_{\nu\eta u} = %\frac{(\nu - \xi\eta)/\sqrt{1-\xi^2} - \Sigma\, (x - \gamma\eta)/\sqrt{1-\gamma^2}}{\sqrt{1 - \Sigma^2}}
  \frac{\nu - \xi\eta - \Sigma u}{\sqrt{2(1 - \xi^2-\Sigma^2)}}\,.
\label{mu}
\end{equation}
In particular, if $\Sigma\simeq 0$ then $p(\delta\ge\delta_c,s|{\rm up}\ S)$ reduces to the simpler form given by \eqn{badApprox}. Moreover, all $\nu$, $\xi$ and $\Sigma$ vanish when $s\to\infty$; then in this limit $\mathrm{erfc}(\mu_{\nu\eta u})\to 1$, which confirms that the approximation for $f(s)$ that one gets from \eqn{approxbacksub} is normalized to unity. 

In the case of a power-law $P(k)$ with a Gaussian filter one has
\begin{equation}
 \frac{\Sigma }{\Gamma\xi} = \frac{1-(r/R)^2}{1+(r/R)^2}\,,
\end{equation}
where $r$ and $R$ are the smoothing scales associated with $s$ and $S$.  It is a simple matter to check that $\Sigma/\sqrt{1-\xi^2}\to 1 - \epsilon^2(1+\epsilon)(1+\Gamma^2)/16$ as $\epsilon\equiv 1-(r/R)^2\to 0$.  (Use of this limiting case helps numerical stability in the $r\to R$ limit.)  

%The $n=-1$ case for constant barriers is again simpler: in this case $\Gamma=1$, $\Sigma/\sqrt{1-\xi^2} = \xi$ and 
%\begin{equation}
% \mu_{\nu\beta x} = \frac{-\nu - \xi\, (x - \gamma\eta)/\sqrt{1-\gamma^2}}{\sqrt{1 - \xi^2}} \qquad (n=-1).
%\end{equation}

Figure~\ref{bsall} compares our back-substitution estimate of $sf(s)$ with that based on \eqn{badApprox}, and with the original approximation of \cite{ms12} for walks associated with a range of power-law power spectra $P(k)\propto k^n$ crossing a barrier of constant height $\delta_c$.  Symbols show the `exact' result obtained by Monte-Carlo methods \citep[following][]{bcek91}.  (Details of our implementation of the back-substitution algorithm are provided in Appendix~\ref{algorithm}.)  We have chosen $n=+1$ and $-2$ to compare directly with the results in \cite{bcek91}, and $n=-1.2$ (rather than $n=-1$) to compare with \cite{ms12}. 

Our Monte-Carlos are in excellent agreement with those shown in Figure~9 of \cite{bcek91}.  E.g., for $n=(+1,-1,-2)$, their Monte-Carlo's cross the curve for walks with uncorrelated steps at $\log_{10}(\delta_c^2/s) = (-1.6, -1.3, -1.1)$.
The approach based on the back-substitution in \eqn{approxbacksub} is now in excellent quantitative agreement with the actual first crossing distribution obtained by direct Monte-Carlo simulation of the (Gaussian smoothed) walks at all $s$. In particular, our approach brings a small correction to $f_{\mathrm{MS}}$, but one which cures the divergence at small $\nu$.  On the scales shown, the correction to $f_\mathrm{MS}$ is significant for $n=-2$ and $-1$, but still small for $n=+1$.  And $f_\mathrm{MS}$ itself is a smaller correction to the completely correlated limit as $n$ increases.  Both these are consistent with the expectation that $\Gamma\nu\sim 1$ sets the new scale.  

%\begin{equation}
% p(\delta\ge\delta_c,s|{\rm first}\ S) \approx 
% \frac{\int_{-\infty}^\mu{\rm d}y\,p(y|\eta)\,{\rm erfc}(\mu_{\nu\eta y}/\sqrt{2})/2}{\int_{-\infty}^\mu{\rm d}y\,p(y|\eta)}
%\end{equation}

Figure~\ref{bsall} shows that $f\sim f_\mathrm{MS}$ at least down to $\Gamma\nu\sim 1$.  This suggests that analytic progress may be made writing $f=f_\mathrm{MS} + (f-f_\mathrm{MS})$ in \eqn{approxbacksub}.  Differentiating with respect to $s$ and rearranging the terms yields 
\begin{align}
  f(s) &= f_\mathrm{MS}(s)
  + \frac{\dd}{\dd s} \int_{\delta_c}^\infty \!\!\dd\delta
  \Big[p(\delta,s) - p_\times^\mathrm{MS}(\delta,s)\Big]\notag\\
    & \quad + \int_0^s \dd S \left[f(S)-f_\mathrm{MS}(S)\right]
              \frac{\dd\,p(\ge \delta_c,s|\mathrm{up~}S)}{\dd s}\,,
\label{bsAnalytic}
\end{align}
where%For what follows, it will also be convenient to define 
\begin{equation}
 \label{pMSs}
  p_\times^\mathrm{MS}(\delta,s)\equiv \int_0^s \dd S\,
  f_\mathrm{MS}(S)\,p(\delta,s|\mathrm{up~}S)
%  p_\mathrm{MS}(\delta\ge\delta_c,s|S).
\end{equation}
corresponds to \eqn{pcross} if one drops the requirement that the walk never crossed before $S$, and thus at small $s$ provides a good approximation to $p(\delta,s)=p_{\times\!}(\delta,s)$ for all $\delta>\delta_c$. Note however that, unlike \eqn{approxbacksub}, this formulation is not valid for uncorrelated steps, since in this case $p(\delta\geq\delta_c,s|\mathrm{up~}S)$ with $S=s$ does not equal 1 but 1/2.

The first term in \eqn{bsAnalytic} provides a correction to $f_\mathrm{MS}$ that depends on the deviation of $p_\times^\mathrm{MS}(\ge \delta_c,s)$ from $p(\ge \delta_c,s)$, which is small at small $s$. The last term contains the correction to $f_\mathrm{MS}$ integrated over $S$, and therefore is expected to kick in only after the first one becomes relevant, as a second order correction.
However, this term is crucial to preserve the normalization of the solution, which being a constraint on the integral of $f$, is non-perturbative in $s$. Inserting \eqn{msApprox} in the back substitution expression (equation~\ref{backsub}) allows one to go beyond the first order correction, keeping the approximation under control also in the large-$s$ regime.
We will discuss an approximation that boils down to ignoring this last term in Section \ref{sec:other}.  For now, we note that our full back-substitution expression is accurate enough to be considered the solution to the original excursion set problem posed by \cite{bcek91}, of walks with correlated steps crossing a threshold of constant height.

\subsection{Moving barriers}
Although \eqn{backsub} did not make any assumption about the barrier shape, the approximations we discussed and the examples we have shown so far all assumed a barrier of constant height.  However, dealing with moving barriers -- barriers that depend on $s$ -- is straightforward. The appropriate generalization of \eqn{msApprox} is
\begin{equation}
 p(\delta\ge b(s)|{\rm up}\ S) \equiv \frac{\int_{b}^\infty{\rm d}\delta 
       \int_{B'}^\infty {\rm d}V (V-B')\,p(\delta,V|B)}
      {\int_{B'}^\infty {\rm d}V (V-B')\, p(V|B)}\,,
 \label{msMoving}
\end{equation}
which inserted into \eqn{backsub}, with $b = b(s)$ on the left hand side, allows solving for $f$. Computing this expression adds no further technical difficulty; in fact, it yields exactly the same result as \eqn{explicit}, upon redefining $X\to -2\Gamma S (B/\sqrt{S})'$. The result is a correction to $f_\mathrm{MS}$ that cures the divergence at very small $s$, if present. Nonetheless, as we now argue, for many moving barriers of current interest $f_\mathrm{MS}$ may already be sufficiently accurate.

%Before we show some examples, it is worth noting that the lower limits of the integrals in \eqn{msMoving} indicate that there are three classes of barriers:  those for which $b/\sqrt{s}$ decreases monotonically as $s$ increases; those for which $b/\sqrt{s}$ has a minimum, after which it increases, and those for which $\dd (b/\sqrt{s})/\dd s$ increases at large $s$.  It may help to think of barriers of the form $\delta_c + \alpha\,s^\omega$, for which the limiting case of the first class has $\alpha>0$ and $\omega=1/2$.  For such barriers the first crossing distribution will be normalized to unity because the $s\to\infty$ limits on both sides of \eqn{normalized} tend to the same value for all $\delta_c$ and $\alpha$ provided $\omega\le 1/2$.  

To make the discussion easier, it may help to consider barriers of the form $b(s)=\delta_c + \alpha\,s^\omega$, and write
\begin{equation}
  \lim_{s\to\infty}\int_0^s \!\dd S\,f(S)\,
  \frac{p(\delta\ge b(s)|{\rm up}\ S)}{p(\delta\ge b(s))} \approx 1\,,
\end{equation}
which is the generalized version of \eqn{normalized}. Insight on the normalization of the solution can then be obtained by studying the limit for $s\to\infty$ of the ratio $\mathrm{erfc}(\mu_{\nu\eta u})/\mathrm{erfc}(\nu/\sqrt{2})$, with $\mu_{\nu\eta u}$ given by \eqn{mu}.
The denominator tends to 1 for $\omega<1/2$ and to $\mathrm{erfc}(\alpha/\sqrt{2})$ for $\omega=1/2$, while for $\omega>1/2$ the limit is either 2 (if $\alpha<0$) or 0 (if $\alpha>0$). The same is true for the numerator: since both $\xi$ and $\Sigma$ become typically proportional to $\sqrt{S/s}$ when $s\gg S$, the presence of $\eta$ and $u$ (which are numbers of order unity) could be reabsorbed rescaling $\delta_c$, and this does not affect the limit. %When $\nu\to\infty$, for $\alpha>0$ and $\omega>1/2$, expanding the erfc's shows that the ratio remains 1 as long as $\nu\xi\to 0$, which means $\omega<1$. 
Thus, the normalization of $f(s)$ obtained from \eqn{msMoving} is always unit, but for $\alpha>0$ and $\omega> 1/2$, in which case no conclusion can be drawn a priori.

\begin{figure}
 \centering
 \includegraphics[width=0.9\hsize]{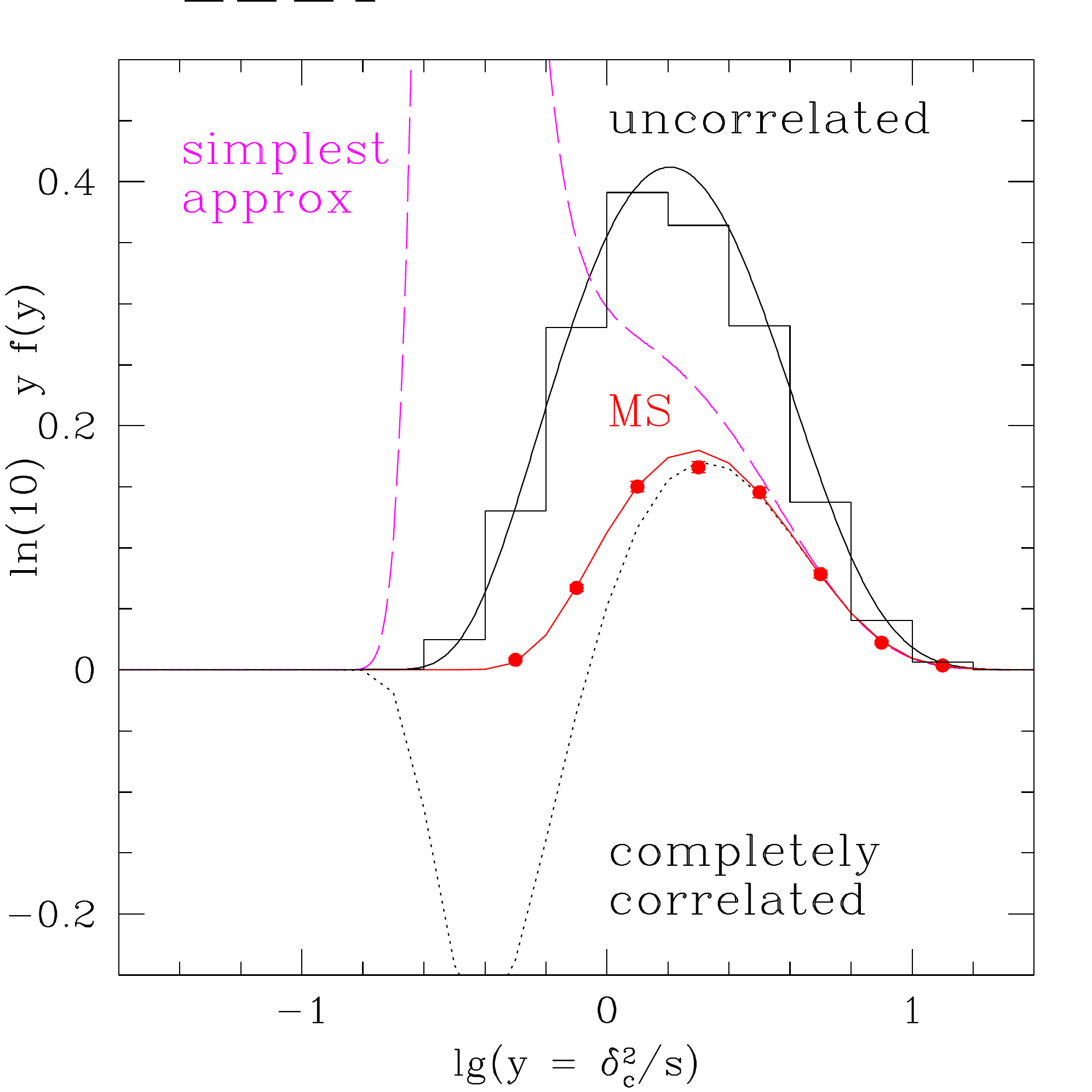}
 \caption{\label{quadratic}
 First crossing distribution $yf(y) = sf(s)$ of a steeply increasing barrier of height $\delta_c[1 + (S/\delta_c^2)^2/4]$, by walks with uncorrelated steps (histogram) and steps having correlations which come from Gaussian smoothing of $P(k)\propto 1/k$ (symbols with error bars).  Upper solid curve shows the result of inserting \eqn{badApprox} in \eqn{backsub} for the uncorrelated steps; which should be exact.  Dashed curve shows the same calculation but now for correlated steps; it lies far from the symbols.  Dotted curve shows $f_\mathrm{PS}$ is also a poor description, since it even goes negative (signalling that there are more walks crossing downwards than upwards).  Lower solid curve shows $f_\mathrm{MS}$; it provides an excellent description of the first crossing distribution.  }
\end{figure}

Regardless of the sign of $\alpha$ (the direction of the barrier), the value $\omega=1/2$ discriminates between barriers that are moving slower or faster than the rms of the distribution. For faster barriers, virtually all crossings must happen when $s\lesssim\delta_c$: at smaller scales diffusion simply cannot keep up with the barrier, and $f(s)$ is suppressed like $\exp(-b^2(s)/s)$. For these barriers, $f_\mathrm{MS}$ is expected to be all that is needed for Cosmology.  On the other hand, sizeable corrections can arise at large $s$ for slower barriers.  We may thus introduce a classification of moving barriers in three categories:
\begin{enumerate}
\item
Fast decreasing barriers ($\alpha<0$ and $\omega>1/2$), whose first crossing distribution is normalized to 1. The probability that $\delta$ lies above $b(s)$ given that it crossed $b(S)$ approaches unity, simply because $b(s)\ll b(S)$ even for $S\lesssim s$.  At small $s$, this makes \eqn{fsgShs} with $g(s)=h(s)=1$ a good approximation (the `completely correlated' limit, which corresponds to setting the term in square brackets in \eqn{sfms} equal to unity). Although this was noted by \cite{pls12}, our \eqn{backsub} is an easy way to see why it works so well. That said, numerical tests show that the full $f_\mathrm{MS}$ is even more accurate, yielding $\int_0^\infty\!\dd s f_\mathrm{MS}(s)\simeq 1$ at less than percent level for linear barriers, and improved accuracy for steeper barriers. Therefore, no further correction is needed.  Examples of such barriers include the excursion set model of the nonlinear probability distribution function \citep{rks98, ls08} or the nonlinear 
 collapse along one axis \citep{sams06} (i.e. excursion set models of large scale filaments). 
\item
Slowly moving barriers ($\omega\leq 1/2$), whose first crossing distribution is also normalized to 1.  In this case, $p(\delta\ge b(s)|{\rm first}\ S)$ is significantly less than 1, making \eqn{fsgShs} a bad approximation (something \citet{pls12} also saw). On the other hand, \eqn{fms} remains a good approximation down to scales of the order of $\delta_c$ and beyond. However, the integral of $f_\mathrm{MS}$ is divergent, signalling that the normalized solution obtained inserting \eqn{msMoving} in \eqn{backsub} provides a relevant correction at very large $s$.
\item
Fast increasing barriers ($\alpha>0$ and $\omega>1/2$). In this case the barrier increases too steeply and not all walks eventually cross, so the normalization factor is less than unity. In fact, this is already true even for $f_\mathrm{MS}$ (which is known to overestimate the result).  Although calculating this factor becomes complicated, the accuracy of our approach is not jeopardized: \eqn{backsub} is always exact, and \eqn{msMoving} is an excellent approximation for $s\gtrsim S$ because -- just as for slowly moving barriers -- much of the error due to the inclusion of walks with earlier crossings cancels out in the ratio.  However, all this brings only minor improvement to $f_\mathrm{MS}$, since most crossings will happen at small $s$, where we already know that $f_\mathrm{MS}$ is accurate enough.  For barriers that increase linearly with $s$, this accuracy was already known \citep{ms12}.  Figure~\ref{quadratic} shows that $f_\mathrm{MS}$ remains accurate even for $\omega
 =2$.
\end{enumerate}

The arguments above are fully general -- they remain true for non-Gaussian processes -- as long as the rms of the distribution grows like $\sqrt{s}$. We conclude therefore that our approach introduces a correction that matters at large $s$, for any mildly scale dependent barrier. This correction guarantees the correct normalization of the solution, and being based on an integral constraint is nonperturbative in $s$. For steeper (increasing or decreasing) barriers, this approach is also correct, but it provides only a small correction to $f_\mathrm{MS}$ before the exponential cutoff of the solution, and is therefore not necessary.

\subsection{Other related approximations}
\label{sec:other}

%Similarly, $p_<(\delta)\equiv p(\delta)-p_{\times\!}(\delta)$ is the probability of reaching $\delta$ without ever crossing $b$. 
An equivalent formulation of the first crossing rate is
\begin{equation}
  f(s) = -\frac{\dd}{\dd s} \int_{-\infty}^{b}\!\!\!
  \dd \delta \Big[p(\delta,s) -p_{\times\!}(\delta,s) \Big]\,,
\label{otherrate}
\end{equation}
with $p_\times(\delta)$ given by \eqn{pcross}.
Using $\int_{-\infty}^b = \int_{-\infty}^{\infty} - \int_b^{\infty}$, and the fact that by construction one has $p_{\times\!}(\delta,s)=p(\delta,s)$ for $\delta>b$, one can in fact check that from this expression one recovers~\eqn{fsexact}.
Note however that here we are looking at $p_{\times\!}(\delta,s)$ for $\delta<b\,$: for barriers of constant height, \eqn{otherrate} is the generalization to correlated steps of the symmetry argument used by \cite{bcek91} for walks with uncorrelated steps (in which case $p_{\times\!}(\delta)$ is a Gaussian with mean $2b$ and variance $s$).  

In the spirit of \citet{ms12}, we can approximate $p_{\times\!}(\delta)$ in \eqn{otherrate} with \eqn{pMSs}, obtaining
\begin{align}
  f(s) \simeq f_\mathrm{PS}(s) +
  \int_0^s\!\!\dd S \!\int_0^\infty \!\!\!\! \dd V \,V 
  \frac{\dd p(\delta\!\leq b, B, V) }{\dd s}\,,
\label{eq:correction}
\end{align}
%\begin{align}
%  f(s) \simeq %f_{\mathrm{MS}}(s) +
%  f_\mathrm{PS}(s) + \frac{\dd}{\dd s}\,
%    p_{\times\!}^{\mathrm{MS}}(\delta\leq b),
%%  \left[p(\delta) -p_{\times\!}(\delta) \right]\,
%\end{align}
where we have %redefined $V\to V+B'$, and 
brought $\dd/\dd s$ inside the integral over $S$ since its action on $s$ in the integration limit gives zero. Evaluating the derivative via the Fokker-Planck equation always involves one derivative with respect to $\delta$, so that the integral over $\delta$ is trivial. The integral over $V$ can thus be computed analytically in full generality, leaving only the integral over $S$ (for which we could find no analytical result) to be computed numerically.

This is apparently an expansion in which $f_\mathrm{PS}(s)$ is the leading order term; however, using again $\int_{-\infty}^b = \int_{-\infty}^{\infty} - \int_b^{\infty}$, the same can be recast as
\begin{align}
 \label{eq:fsapprox}
  f(s) \simeq f_{\mathrm{MS}}(s)
  + \frac{\dd}{\dd s} \int_{b}^{\infty}\!\!\!\! \dd \delta  
  \left[p(\delta,s) - p_{\times\!}^{\mathrm{MS}}(\delta,s)\right],
 %  \left[p(\delta) -p_{\times\!}(\delta) \right]\,,
\end{align}
where now the second term no longer vanishes. Therefore, the deviation of $p_{\times\!}^{\mathrm{MS}}$ from $p(\delta,s)$ provides at small $s$ the first order correction to the approximation $f(s) \simeq f_{\mathrm{MS}}(s)$. This deviation is small in this regime, which is why $f_{\mathrm{MS}}$ works so well. %Being a positive definite integral that includes more trajectories than the exact result of \eqn{fsexact}, $f_{\mathrm{MS}}$ overestimates $f(s)$. Similarly, $p_\times^{\mathrm{MS}}$ contains more trajectories than the exact $p_\times$, so that more walks are removed from $f_{\mathrm{MS}}$ than should be, and thus \eqn{eq:correction} underestimates $f(s)$.  
However, the expression above corresponds to ignoring the final term in \eqn{bsAnalytic}, which as we have discussed becomes important at large $s$.

\begin{figure}
 \centering
 \includegraphics[width=0.9\hsize]{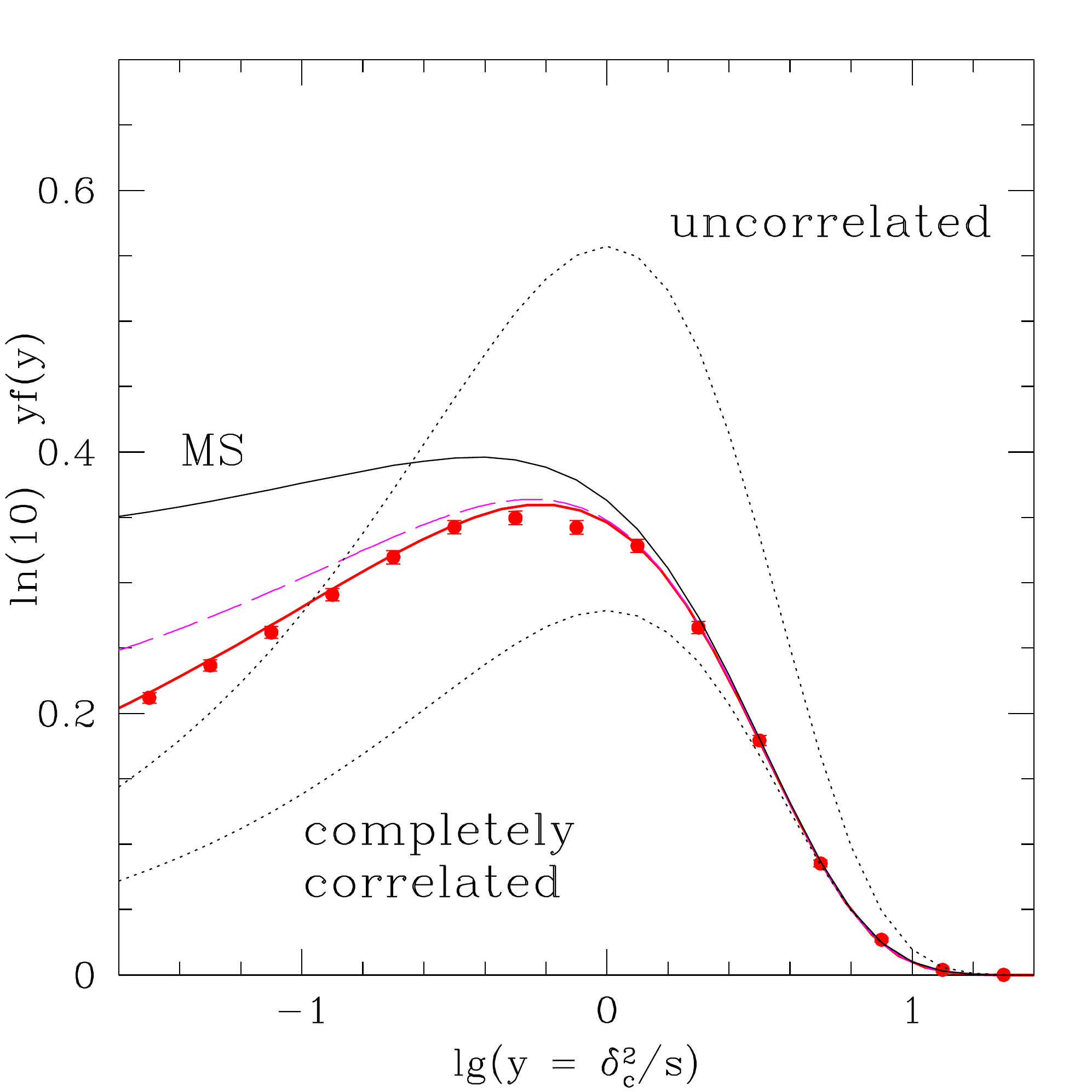}
 \caption{\label{fig:ftoy}
 First crossing distribution $yf(y) = sf(s)$ of a constant barrier of height $\delta_c$, by walks with Markovian velocities which have $\gamma=1/2$.  Filled circles show the Monte-Carlo'd distribution; thick solid curve shows the full back-substitution expression; dashed curve shows \eqn{ftoy}; and thinner solid curve shows the simpler approximation of \citet{ms12}.  The two dotted curves show the approximation of \citet{ps74} and twice this value. 
}
\end{figure}

To illustrate this approximation, we ran walks whose velocities (not walk heights!) are Markovian. These can be tuned, following methods given in Musso \& Sheth (2013b), to mimic to remarkable accuracy the results of the actual excursion set walks for a given power spectrum and filter; yet their Monte-Carlo simulation benefits of the appealing properties of a Markovian process, which makes it easy to reach very large values of $s$.
%, \eqn{eq:correction} for a barrier of constant height $b=B$ can be written exactly:  
The relevant parameters for the back-substitution algorithm are 
\begin{displaymath}
 \label{equivalence}
 \gamma = \frac{1}{2}, \ \ %\Gamma^2 = \frac{1}{3},\ 
    \xi = \sqrt{\frac{S}{s}}\, \frac{3 - S/s}{2} \ \ {\rm and}\ \ 
 \Sigma = \sqrt{3\,\frac{S}{s}}\frac{1-S/s}{2}
\end{displaymath}
for $S\le s$.  
This correlation structure, which is quite different from that for Gaussian smoothing, is not too different from that for TopHat smoothing of $P(k) \propto k^{-2}$, which has $\gamma = 1/\sqrt{6}$ and $\xi = \sqrt{S/s}\, (5 - S/s)/4$.  In addition, $\gamma\sim 1/2$ is a reasonable approximation to TopHat smoothed $\Lambda$CDM \citep[see Figure~4 in][]{pls12}. We compared the results of the Monte-Carlo walks with the full approximate back-substitution solution (equation \ref{msApprox}), and with its first order term (equation \ref{eq:correction}), which for these walks reads
\begin{align}
  sf(s) &\approx sf_\PS(s) + \int_0^1\frac{\dd y}{y} \frac{1}{1-y}
  \int_0^\infty \!\!\!\dd w \, w^2 \, \notag \\
  &\times  \frac{e^{-w^2/2y}}{\sqrt{2\pi y}}
  \frac{e^{-[w-2(b/\sqrt{s})]^2/6y}}{\sqrt{6\pi y}}
  \frac{e^{-w^2/2(1-y)}}{\sqrt{2\pi(1-y)}} \,
\label{ftoy}
\end{align}
(see Musso \& Sheth 2013b for details).  
Figure~\ref{fig:ftoy} shows that \eqn{ftoy}, which boils down to ignoring the final term in \eqn{bsAnalytic}, is a better approximation than is $f_\mathrm{MS}$, and the full back-substitution expression is even more accurate.

An alternative formulation of our integral equation approach can be obtained by replacing the left hand side of \eqn{backsub} with the fraction of walks that are crossing the barrier $b$ upwards, regardless of whether or not they had crossed before: that is, with $f_\mathrm{MS}(s)$.  Then consider a lower barrier $c = b - b_0$, for some constant $b_0>0$.  Since each of the walks that cross $b$ at $s$ must have first crossed $C\equiv c(S)$ at some $S< s$ one could replace \eqn{backsub} with 
\begin{equation}
 f_\mathrm{MS}(s) =
  \int_0^s \dd S\, f(S)\,f_\mathrm{MS}(s|{\rm first}\ S),
 \label{backsub2}
\end{equation}
where $f(S)$ is the first crossing distribution for the lower barrier $c$, and $f_\mathrm{MS}(s|{\rm first}\ S)$ is the fraction of walks that cross the barrier $b$ at $s$ going upwards (whether or not for the first time), and that had crossed the lower barrier $c$ for the first time at $S< s$.  We may then approximate 
\begin{align}
 f_\mathrm{MS}(s|{\rm first}\ S)%& \approx
%   \int_{b'}^\infty \dd v\,v\,p(b,v|C,V\ge C')\nonumber\\
   & \approx \int_{b'}^\infty \!\!\dd v\,v\,p(b,v|C) 
   \equiv f_\mathrm{MS}(s|C)
\end{align}
where the final expression is the conditional distribution of \citet[][see their equations~28 and Figure~2]{mps12}.  This approximation also requires knowledge of only a trivariate distribution, but this time for $(b,v,C)$ rather than $(b,B,V)$, and it too can be solved by back-substitution.  

%To see what it implies, write $f = f-f_\mathrm{MS}+f_\mathrm{MS}$ and then differentiate with respect to $s$.  This yields 
%\begin{align}
%  \frac{\dd f_\mathrm{MS}(s)}{\dd s} & \approx f(s)-f_\mathrm{MS}(s)
%   + \frac{\dd}{\dd s}\int_0^s \dd S\, f_\mathrm{MS}(S)\,
%     f_\mathrm{MS}(s|C)  \nonumber\\
%   & \quad + \int_0^s \dd S\, [f(S)-f_\mathrm{MS}(S)]\,
%      \frac{\dd f_\mathrm{MS}(s|C)}{\dd s}\,, %\nonumber\\
%%   & \approx f(s) - f_\mathrm{MS}(s)
%%       + \frac{\dd}{\dd s}\int_0^s \dd S\, f_\mathrm{MS}(S)\,f_\mathrm{MS}(s|C)\nonumber
%\end{align}
%where, as for the previous approximation, we can also assume that the last term is a second order correction in the small-$s$ regime, where $f\sim f_\mathrm{MS}$.
%Therefore we get
%\begin{equation}
%  f(s) \approx f_\mathrm{MS}(s)
%  + \frac{\dd}{\dd s}\!\left[f_\mathrm{MS}(s) - \!%\frac{\dd}{\dd s}
%  \int_0^s \!\!\dd S\, f_\mathrm{MS}(S)\,f_\mathrm{MS}(s|C)\right]\!.
%\end{equation}
%Since
%$\int_0^s \dd S\, f_\mathrm{MS}(S)\,f_\mathrm{MS}(s|C) \approx f_\mathrm{MS}(s)$
%to leading order, the expression above shows that this approach will indeed have  $f(s) = f_\mathrm{MS}(s)$ plus corrections that only matter at large $s$.  

%%%%%%%%%%%%%%%%%%%%%%%%%%%%%%%%%%%%%%%%%%%%%%%%%%%%%%%%%%%%
\section{Discussion}
%%%%%%%%%%%%%%%%%%%%%%%%%%%%%%%%%%%%%%%%%%%%%%%%%%%%%%%%%%%%

Previous work on the first crossing distribution has shown the power of including the constraint that walks must cross upwards \citep{ms12}.  The simplest version of this approach uses the joint distribution of the walk height and its slope on a given scale, and reduces the problem from one involving an infinite number of correlated variables to only two (equation~\ref{fms}).  This works well down to scales where a large fraction of walks may have zig-zagged their way around the barrier more than once.  \cite{ms13a} have shown how to account for walks with more zigs and zags order by order, but each order in this expansion requires the introduction of two additional (correlated) variables. Already the first correction contains integrals that cannot be done analytically.  

Our approach here is different:  we started from an integral equation, \eqn{backsub}, and showed how to use the same upcrossing requirement  to account for the effects of walks with multiple-crossings (equation~\ref{msApprox}). Although the solution still cannot be computed analytically, it requires knowledge of just one additional variable. Overall, the approach we presented uses the joint distribution of three variables:  the walk height at the scale of interest, and its height and slope on a larger scale.  (Along the way, we showed that not including the slope is a bad approximation.)  The resulting first crossing distribution, obtained by back substitution, is only an approximation to the exact solution, but an extremely good one (Figure~\ref{bsall}), and is guaranteed to be correctly normalized: it integrates to one for all barriers that do not grow too fast with scale (for very steep barriers the normalization is less than one, as expected). 

Over the range of scales that is typically of interest in Cosmology -- the large mass regime -- our numerical solution yields a small correction to \eqn{fms}, the simple result of \citet{ms12} for walks with correlated steps.  However, the normalization constraint is non-perturbative in $s$ and keeps the solution under control even in the very large $s$ regime (i.e.~ for very small masses), which we have checked against Monte-Carlo simulation of the walks (Figure~\ref{bsall}). This remains true for the walks generated by those filters and power spectra for which \eqn{fms} performs less well (as $n$ becomes more negative), and even for walks with uncorrelated steps, for which the present approach returns the exact result, while \eqn{fms} diverges.

Although we illustrated our results using walks from Gaussian smoothing of a Gaussian field with power-law power spectrum crossing a barrier of constant height, we found excellent agreement for walks with a rather different correlation structure (Figure~\ref{fig:ftoy}), which mimics the TopHat smoothing of a $\Lambda$CDM power spectrum.  In this respect, we feel we have demonstrated that our approach has effectively solved the original excursion set problem posed by \cite{bcek91}:  that of walks with correlated steps crossing a barrier of constant height.  And it has done so by making use of just three correlated variables. We believe that this spectacular agreement stems from applying the upcrossing approximation of \citet{ms12} to numerator and denominator of \eqn{msApprox}, with the error that appears at small masses in both largely cancelling out from their ratio. 

In addition, we argued that our formulation \citep[like that of][]{ms13a} applies as well to moving barriers (simply replace equation~\ref{msApprox} with equation~\ref{msMoving}) and to non-Gaussian walks.  However, we also argued that for steeply increasing barriers $f_\mathrm{MS}$ is in fact already accurate enough (Figure~\ref{quadratic}), and the same is true for steeply decreasing barriers (in which case even the simpler $f_\mathrm{PS}$ does well at large scales).  I.e., most of the complications here only matter in the intermediate regime in which the barrier to be crossed is a weaker function of smoothing scale $s$ than is the rms of the underlying distribution.  

Solution of the integral equation is trivial and fast, so although we provided an analysis of the leading order corrections to the Musso-Sheth formula that it yields (Figure~\ref{fig:ftoy} indicates these are indeed more accurate, though not as accurate as the full solution of course), and we provided an example of another way in which one might have set-up the back-substitution calculation (equation~\ref{backsub2}), we have not bothered to search for an analytic approximation to the full solution.  In part, this is because we believe that the Musso-Sheth formula, our \eqn{fms}, is sufficiently accurate for most applications of cosmological interest.

There is however the additional concern that the fundamental assumption that equates the first crossing distribution to halo statistics, \eqn{ansatz}, is not accurate \citep{aj90, cusp98, n01}, and that a better treatment of the problem is to select a subset of the walks that have steeper slopes \citep{ms12, ps12}.  In effect, this results in an additional weighting term on the ensemble of walks \citep{ms12, cs13} that the analysis here does not include.  We expect this weight to simply enter as an additional factor of $f(V)/(2\pi R^2)^{3/2}$ in both the numerator and denominator of \eqn{msMoving}, but we have not checked this.  If this is indeed all there is to it, then because this additional weight selects steeper walks, its inclusion will matter less than when this weight is not included, since the requirement of steeper walks itself fixes some of the divergence at large $s$:  i.e., any correction to \eqn{msApprox} (once the additional weight on the distribution of $v$ ha
 s been included) will be small.  

%%%%%%%%%%%%%%%%%%%%%%%%%%%%%%%%%%%%%%%%%%%%%%%%%%%%%%%%%%%%
\section*{Acknowledgments}
The work of MM is supported by the ESA Belgian Federal PRODEX Grant No.~4000103071 and the Wallonia-Brussels Federation grant ARC No.~11/15-040.  RKS is supported in part by NSF-AST 0908241.

\bibliography{mybib}{}

\appendix 
\section{Gaussian smoothing of a power-law spectrum}\label{n=-1}

We remarked in the main text that Gaussian smoothing of $P(k)\propto k^{-1}$ allows for some clean analytic results.  These are collected in this Appendix.  

We first consider the simple approximation, \eqn{badApprox}.  For Gaussian smoothing of a Gaussian field with $P(k)\propto k^n$ with $n=-1$, the cross-correlation of the field at two different scales is 
\be 
 \frac{S_\times}{S} \equiv \frac{\avg{\!\delta\Delta\!}}{\avg{\!\Delta^2}\!} 
                       = \frac{2}{1 + S/s}.
\ee
This makes the term in square brackets of \eqn{badApproxGauss} particularly simple:  
\be
 \label{minusnu}
 \frac{\delta_c(1 - S_\times/S)}{\sqrt{s - S_\times^2/S}}
  = -\frac{\delta_c}{\sqrt{s}}.
\ee  
Since it is independent of $S$, 
 $p(\delta\ge\delta_c,s|\delta_c,S)$ can be written as $g(S)h(s)$, with
 $g(S)=1$ and $h(s) = {\rm erfc}(-\delta_c/\sqrt{2s})/2$.  
Hence, this is the first non-trivial case in which equation~(\ref{fsgShs}) can be used, yielding 
%It can, therefore be taken out of the integral.  Moving it to the left hand side and differentiating with respect to $s$ then yields
\begin{equation}
 sf(s) = \frac{{\rm d}}{{\rm d}\ln s}
        \frac{{\rm erfc}(\delta_c/\sqrt{2s})}{{\rm erfc}(-\delta_c/\sqrt{2s})}
 %       \nonumber\\
 %      &=& \frac{\delta_c\,\exp(-\delta_c^2/2s)}{2^{1/2}{\rm erfc}(-\delta_c/\sqrt{2s})\sqrt{2\pi s}}[1 - \frac{{\rm erfc}(\delta_c/\sqrt{2s})}{{\rm erfc}(-\delta_c/\sqrt{2s})}]\nonumber\\
 %      &=& \frac{\delta_c\,\exp(-\delta_c^2/2s)}{2^{1/2}{\rm erfc}(-\delta_c/\sqrt{2s})\sqrt{2\pi s}} \frac{2{\rm erf}(\delta_c/\sqrt{2s})}{{\rm erfc}(-\delta_c/\sqrt{2s})}]
 %      = \frac{\nu\,\exp(-\nu^2/2)/\sqrt{2\pi}}
 %             {[1 + {\rm erf}(\nu/\sqrt{2})]^2/2}.
       = \frac{sf_{\rm PS}(s)}
              {[1 + {\rm erf}(\nu/\sqrt{2})]^2/4}\,.
 \label{nminus1}
\end{equation}

At $\nu\gg 1$ this result correctly tends to $sf_{\rm PS}(s)$, but the leading order multiplicative correction factor of $[1 + (2/\nu)\, {\rm e}^{-\nu^2/2}/\sqrt{2\pi}]$ is larger than for $sf_{\rm MS}(s)$, for which the correction is $1 + {\rm e}^{-\Gamma^2\nu^2/2}/\sqrt{2\pi}/(\Gamma\nu)^3$ (remember $\Gamma^2=(n+3)/2$ for Gaussian smoothing, and we are considering $n=-1$, for which $\Gamma^2=1$).   Since $f_\mathrm{MS}$ is rather close to the correct answer, this indicates that approximation~(\ref{badApprox}) predicts more crossings at large $s$ than actually occur:  it is not a very good approximation for the special case of $n=-1$.  Since there is nothing particularly special about $n=-1$, other than that it could be solved analytically, we expect \eqn{badApprox} to remain a bad approximation in general.  

The poor performance of \eqn{nminus1} is confirmed by the results shown in Figure~\ref{bsall} of the main text (the Figure actually shows results for $n=-1.2$, but $n=-1$ is very similar). This approximation is therefore not a useful correction to $f_\mathrm{MS}$ (it actually makes it worse). Still, it provides a nontrivial fully analytic solution to \eqn{backsub}, and is thus a very useful test to check the accuracy of the numerical back-substitution algorithm that we adopt to solve also the more general case. We used it for this purpose when developing the code.  

For linear barriers, $b(s) = \delta_c\,(1 + \alpha\, s/\delta_c^2)$, the term in square brackets of \eqn{badApproxGauss} becomes 
\be
 \label{erfcterm}
 \frac{b(s) - (S_\times/S)\, B(S)}{\sqrt{s - S_\times^2/S}}
 %= \frac{\delta_c\,(1 - S_\times/S) + (\beta/\delta_c) (s - S_\times)}{\sqrt{s - S_\times^2/S}}
 %= \frac{\delta_c\,(1 - S_\times/S) + (\beta/\delta_c) (s+S - 2S)/(1 + S/s)}{\sqrt{s - S_\times^2/S}}
 %= \frac{\delta_c\,(1 - S_\times/S) + (1-S_\times/S)(\beta/\delta_c)\, (s - S)/(1 + S/s)/(1-Sx/S))}{\sqrt{s - S_\times^2/S}}
 %= \frac{\delta_c\,(1 - S_\times/S) + (1-S_\times/S)(\beta/\delta_c)\, (s - S)s/(S-s)}{\sqrt{s - S_\times^2/S}}
 %= \frac{\delta_c\,(1 - S_\times/S) - (1-S_\times/S)(\beta/\delta_c)\,s}{\sqrt{s - S_\times^2/S}}
 = -\frac{\delta_c}{\sqrt{s}} + \alpha\,\frac{\sqrt{s}}{\delta_c},
\ee
% note that  db/sqrts /ds = d/ds (dc/sqrt(s) + beta/dc sqrt(s))
%                         = [-dc/sqrt(s) + beta sqrt(s)/dc]/2/s
which is still independent of $S$.  This means that $sf(s)$ can be obtained analytically.  If we define $\beta(s)\equiv b/\sqrt{s}$ then the term above is $2\,\dd \beta/\dd\ln s = 2s\beta'$, and so 
\be
 sf(s) = sf_\mathrm{PS}(s)\,
         \frac{{\rm e}^{2\alpha}\, \beta\, {\rm erfc}(\beta/\sqrt{2}) - 
  (2s\beta'){\rm erfc}(\sqrt{2}s\beta')} 
              {-s\beta'\,[{\rm erfc}(\sqrt{2}s\beta')]^2}.
\ee
When $\alpha=0$ then $\beta=\nu$ and $\dd\beta/\dd\ln s = -\nu/2$ so this reduces to \eqn{nminus1}.  
%\be
% sf(s) = sf_\mathrm{PS}(s)
%         \frac{{\rm erfc}(\nu/\sqrt{2}) + {\rm erfc}(-\nu/\sqrt{2})} 
%              {1/2[{\rm erfc}(-\nu/\sqrt{2})]^2}
%       = sf_\mathrm{PS}(s)\frac{4}{[{\rm erfc}(-\nu/\sqrt{2})/2]^2}
%\ee

More insight into the general case comes not so much from this expression for $sf(s)$, as from \eqn{erfcterm}.  For $\beta<0$ (linearly decreasing barriers) this will make the ${\rm erfc}$ of \eqn{badApproxGauss} tend to unity -- the value associated with completely correlated steps -- at both large and small $s$.  This is an explicit demonstration of the statement in the main text about the accuracy of the completely correlated limit for barriers which decrease with $s$.  On the other hand, $\beta > 0$ will have ${\rm erfc}\to 0$ for large $s$, which is why \eqn{badApprox} does not work well for increasing barriers.  

We now move on to consider the approximation that fared substantially better, \eqn{msApprox}.  We remarked that there too, at least in some regimes, the full expression could be reduced to the form of \eqn{gShs}.  This almost happens for Gaussian smoothing of $n=-1$.  To see this, of the integrals over $V$ and $\delta$ which \eqn{msApprox} requires, do the one over $V$ first.  This will yield $p(\delta|B)$ times $\avg{\!V|B,\delta\!}$ times the term in square brackets in \eqn{fms}, but with $X\to \avg{\!V|B,\delta\!}/\sigma_{V|B,\delta}$.
%\be
% X\to \frac{\avg{\!V|B,\delta\!}}{\sigma_{V|B,\delta}}
%\ee
% int dV V p(V|B) = sigV int dx x p(x|nu)
%    = sigV int dx (x-<x|nu>+<x|nu>) p(x|nu)
%    = sigV <x|nu> [p(x>0|nu) + sigx|nu/<x|nu> int_(-<x|nu>/sig) dy y p(y)
%    = sigV <x|nu> [p(x>0|nu) + sigx|nu/<x|nu> exp(-<x|nu>^2/sig^2/2)/sqrt(2pi)
%    = sigV g nu [p(x>0|nu) + sigx|nu/<x|nu> exp(-<x|nu>^2/sig^2/2)/sqrt(2pi)
%
%    = <V|B,d> [p(x>0|nu,B) + sigx|nu/<x|nu> exp(-<x|nu>^2/sig^2/2)/sqrt(2pi)
% 
Now,
\be
 \frac{\avg{\!V|B,\delta\!}}{\sigma_V} 
          %= g \eta + (<Vd> - g \xi)/(1-\xi^2) (\nu - \xi\eta)
           = \gamma\eta + \frac{\sqrt(1-\gamma^2)\Sigma}{\sqrt1-\xi^2}
                          \frac{\nu - \xi\eta}{\sqrt1-\xi^2}
\ee
and 
\be
 \sigma^2_{V|B,\delta} = \sigma^2_V\,(1 - \gamma^2)\,
                       (1 - \xi^2 - \Sigma^2)/(1 - \xi^2).
\ee
When $n=-1$, then $\Gamma =1$ and $\Sigma/\sqrt{1-\xi^2} = \xi$, so 
\be
 X = %\frac{\avg{\!x|\eta,\nu\!}}{\sigma_{x|\eta,\nu}} = 
  \frac{\eta + \xi\,(\nu - \xi\eta)/\sqrt{1-\xi^2}}{\sqrt{1 - \xi^2}},
\ee
where we should think of $\eta$ as $\avg{\!V|B}/\sigma_{V|B}$. 
%and
%\be
% \frac{\avg{\!x|\eta}}{\sigma_{x|\eta}} = \Gamma\eta = \eta.
%\ee
Because the integral which remains to be done is over $\delta\ge \delta_c$, we know that 
\be
 X \ge \frac{\eta + \xi\,(\nu_c - \xi\eta)/\sqrt{1-\xi^2}}{\sqrt{1 - \xi^2}}
 = \frac{\eta - \xi\nu_c}{\sqrt{1 - \xi^2}} = \eta
\ee
where we used \eqn{minusnu} to set $(\nu_c - \xi\eta)/\sqrt{1-\xi^2} = \nu_c$, 
%because 
%$\frac{\nu_c - \xi\eta}{\sqrt1-\xi^2} 
%   %= \nu_c(1 - \xi\sqrt(s/S))/\sqrt(1-\xi^2)
%    = -\nu_c$.
and we then simplified $\eta - \xi\nu_c$ similarly.  
%But this means 
% $X \ge \frac{\eta(1 - S/s)/(1+S/s)}{(1-S/s)/(1+S/s)} = \eta$, 
That $X\ge\eta = \avg{\!V|B}/\sigma_{V|B}$ means the term in square brackets in \eqn{fms} is unity for a greater range of $B$ than it is for just $\avg{\!V|B\!}/\sigma_V$.  When it is unity, then integrating $\avg{\!V|B,\delta\!}$ over $\delta\ge\delta_c$ yields 
%$\sqrt{1-\gamma^2}$ times 
%$\eta\,{\rm erfc}(-\nu_c/\sqrt{2})/2 + \xi\,{\rm e}^{-\nu_c^2/2}/\sqrt{2\pi}$
  % <V|B> = sigV g eta = sigV sqrt(1-g^2) eta
$\avg{\!V|B\!}$ times ${\rm erfc}(-\nu_c/\sqrt{2})/2 + (\xi\nu_c/\eta)\,{\rm e}^{-\nu_c^2/2}/\sqrt{2\pi}\nu_c$.  

Notice that this only depends on $S$ because of $\xi \nu_c/\eta = 2(S/s)/(1 + S/s)$.  In the limit where $\xi\nu/\eta\to 0$, this term will be the same as for our simplest approximation.  As a result, the expression for $f$ from this approximation will be \eqn{nminus1}, with $f_\mathrm{PS}$ replaced by $f_\mathrm{MS}$.  Since the net effect of this term will be to provide a correction to $f_\mathrm{MS}$ rather than to $f_\mathrm{PS}$, and the correction to $f_\mathrm{PS}$ is already too large, we know this will overestimate the true $f$.  

The other limit, $\xi\nu/\eta\to 1$, is more interesting, since then this term becomes $f_\mathrm{MS}(s)/f_\mathrm{PS}(s)$ which is independent of $S$.  This separability makes 
\be
 sf(s) = sf_\mathrm{MS}(s)\,\frac{1 + {\rm erfc}(\nu/\sqrt{2})/2\nu^2}
                                {[f_\mathrm{MS}(s)/f_\mathrm{PS}(s)]^2}\,;
\ee
the term that multiplies $sf_\mathrm{MS}(s)$ is unity at $\nu\gg 1$, and gently decreases to a minimum of 0.9 at $\nu\sim 1$ before the approximation breaks down and this term starts to exceed unity again at $\nu<0.8$.  This is the 
$sf(s)\approx sf_\mathrm{MS}(s)$ behaviour we promised in the main text.  

We could have gone further by setting $\xi \nu_c/\eta\to S/s$, which has the right limiting behaviour at both large and small $S/s$.  The result is messier, as we must now take two derivatives with respect to $s$ to get an analytic expression for $f$, but the result is still analytic.  Clearly, one could expand $\xi \nu_c/\eta$ in powers of $S/s$, each one of which will require another derivative with respect to $s$ to obtain $f$.  We have not pursued this further, since we are in any case assuming that the term in square brackets in \eqn{fms} is unity, and this will break down as $s$ increases.  One could also imagine expanding this term in powers of $X$ when integrating over $\delta\ge\delta_c$ but we have not done so because the other approximation we provide in the main text (e.g. equation~\ref{bsAnalytic}) is more efficient and general.  

\section{Back-substitution}\label{algorithm}
We solve \eqn{backsub} numerically as follows.  The first step is to write the integral on the right hand side as a discrete sum:
\be
 p_j = \sum_{i\le j} F_i\,P_{ji},
\ee
where we have absorbed the step size into the definition of $F_i$.  
Recall that $p$ and $P$ are known, and the problem is to find $F$.  The logic of the solution is set by noting that $p_1 = F_1\,P_{11}$, yields an expression for $F_1$, which can be used in the expression for $p_2 = F_1\,P_{12} + F_2\,P_{22}$ to yield an expression for $F_2$, and so on.  The general case reads:  
\be
 F_j = \frac{p_j - \sum_{i\le j} F_i\,P_{ji}}{P_{jj}}.
\ee
Implementing this is particularly simple in languages that have been optimized for matrix manipulations.  As a result, most of the hard work is in precomputing the matrix elements $P_{ji}$, since these require numerical integrals (e.g. equation~\ref{msApprox}).

The only remaining questions are the step size and the range in $S$ over which to step.  Since we know the solution will have exponential tails, it makes sense to take steps that are spaced equally in logarithimic rather than linear intervals.  Moreover, the speed of the algorithm is set by the number of steps, so using log-steps yields significant speed up, because it allows us to cover a large dynamical range in $S$ with few steps.  In effect, this means that we treat $F_i$ in the expression above as being $(\Delta\ln S)\,Sf(S)$, for some constant $\Delta\ln S$.  I.e., the quantity we care about, $Sf(S)$, is obtained by dividing the $F_i$ by the step size $\Delta\ln S$.  The accuracy of the algorithm depends on this step size:  tests with the known analytic solutions discussed in the main text have shown that $\Delta\ln S\le 0.1$ yields sufficient accuracy.  The results in the main text were obtained with $\ln (S/\delta_c^2)$ running over $[-5,5]$, and $\Delta\ln S = 0.1$. 
  This required 100 steps, which took about one second on a standard laptop; neither one of us is an accomplished programmer.

\label{lastpage}

\end{document}